
\input jnl
\input reforder
\input defs

\title Correlation Functions in Disordered Systems

\author E. Br\'ezin

\affil
 Laboratoire de Physique Th\'eorique
\'Ecole Normale Sup\'erieure
24 Rue Lhomond
75231 Paris, France

\author A. Zee

\affil
Laboratoire de Physique Th\'eorique
\'Ecole Normale Sup\'erieure
24 Rue Lhomond
75231 Paris, France
and
Institute for Theoretical Physics
University of California
Santa Barbara, CA 93106-4030 USA

\abstract
{Recently, we found that the correlation between the eigenvalues of random
hermitean matrices exhibits universal behavior. Here we study this universal
behavior and develop a diagrammatic approach which enables us to extend our
previous work to the case in which the random matrix evolves in time or
varies  as some
external parameters vary. We compute the current-current correlation function,
discuss various generalizations, and compare our work with the
work of other authors. We study the distribution of eigenvalues of Hamiltonians
consisting of a sum of a deterministic term and a random term. The correlation
between the eigenvalues when the deterministic term is varied is calculated.}

\head{I.   Introduction}

We have been studying correlations  between energy eigenvalues in random matrix
theory\refto{WIG,POR,MEH,JER} in an attempt to uncover possible universal
behavior in disordered systems. Let us begin by summarizing the main results of
our previous work.\refto{BZ1,BZ2}
Consider an ensemble of $N$ by $N$ hermitean matrices $\varphi$ defined by the
probability distribution
$$
P(\varphi)={1\over Z}e^{-N^2\cH(\varphi)} \eqno(1.1)
$$
with
$$
\cH(\varphi)={1\over N}Tr V(\varphi) \eqno(1.2)
$$
for any even polynomial $V$ and with $Z$ fixed by the normalization $\int d
\varphi
P(\varphi)=1$.
Define the Green's functions
$$
\eqalignno{
G(z)&\equiv\vev{{1\over N}tr{1\over z-\varphi}}&(1.3)\cr
G(z,w)&\equiv\vev{{1\over N}tr{1\over z-\varphi}{1\over N}tr{1\over
w-\varphi}}&
(1.4)
\cr}
$$
and so forth, where
$$
\vev{0 (\varphi)} \equiv \int d\varphi0(\varphi)P(\varphi)\eqno(1.5)
$$
The density of eigenvalues is then given by
$$
\rho(\mu)=\vev{{1\over N}tr\de(\mu-\varphi)}={-1\over\pi}Im G(\mu+i\eps)
\eqno(1.6)
$$
and the correlation between eigenvalues, by
$$
\eqalign{
\rho(\mu,\nu)&=\vev{{1\over N}tr\de(\mu-\varphi){1\over
N}tr\de(\nu-\varphi)}\cr
&=(-1 /4\pi^2)(G(++)+G(--)-G(+-)-G(-+))\cr}\eqno(1.7)
$$
with the obvious notation
$$
G(\pm,\pm)\equiv G(\mu\pm i\eps,\nu\pm i\de)\eqno(1.8)
$$
(signs uncorrelated).

In the large $N$ limit, $\rho(\mu,\nu)\to\rho(\mu)\rho (\nu)$, and thus it is
customary to define the connected correlation
$$
\rho_c (\mu,\nu)\equiv\rho(\mu,\nu)-\rho(\mu)\rho(\nu) \eqno(1.9)
$$
Note that the factors of $N$ are chosen in our definitions such that the
interval $[-a,+a]$ over which $\rho(\mu)$ is non-zero is finite (\ie of order
$N^0$) in the large $N$ limit.

For applications to disordered system $\varphi$ is often thought of as the
Hamiltonian. Its eigenvalues then describe the energy levels of the system. In
some applications, $\varphi$ is related to the transmission matrix.\refto{PICH}
The density of and correlation between its eigenvalues tell us about the
conductance fluctuation in disordered metals.

The density of eigenvalues has long been known in the literature\refto{BIPZ} to
have the form
$$
\rho(\mu)={1\over\pi}P(\mu)\sqrt{a^2-\mu^2}   \eqno(1.10)
$$
where $P(\mu)$ is a polynomial of degree $2p - 2$ if the potential $V$ is of
degree $2p$. The polynomial $P(\mu)$ and the endpoint of the spectrum $a$
depend on $V$ in a complicated way. In our recent work, we have focussed on the
correlation between eigenvalues.

We have obtained the following results,\refto{BZ1,BZ2} all in the large $N$
limit.

\item{(1.)} Using the method of orthogonal polynomials, we determine
$$
\eqalign{
\rho_c(\mu,\nu)& =-({a\over4N})^2(\mu-\nu)^{-2} [f(\mu)f(\nu)]^{-1}\cr
&\times\Bigg\{[\cos[\varphi(\mu)]-\cos[\varphi(\nu)]] [\cos[N(h(\mu)+(\nu))]
+\cos[N(h(\mu)-h(\nu))]]\cr
&+[\sin[\varphi(\mu)]-\sin[\varphi(\nu)]]\sin[N(h(\mu)+h(\nu))]\cr
&+[\sin[\varphi(\mu)]+\sin[\varphi(\nu)]] \sin[N(h(\mu)-h(\nu))]\Bigg\}^2.\cr}
 \eqno(1.11)
$$
The various functions $f(\mu),\varphi(\mu)$ and
$h(\mu)$  are given in \Ref{BZ1} and depend on $V$. Thus, this rather
complicated expression depends on $V$ in detail.

\item{(2.)} With $\mu-\nu$ close together, of order $N^{-1}$  times some number
large compared to unity
so that there can still be a finite number of eigenvalues between $\mu$ and
$\nu$, and with both $\mu$ and
$\nu$ at a finite distance from the end points of the spectrum, we
obtain the universal result
$$
K(\mu,\,\nu) \rightarrow
{\sin[2\pi N\de\mu\rho(\bar\mu)] \over2\pi N\de\mu}, \eqno(1.12)
$$
where $\bar\mu\equiv{1\over2}(\mu-\nu)$ and $\de\mu\equiv{1\over2}(\mu-\nu)$.
This means that the correlation function takes the form
$$
\rho(\mu,\nu)=\rho(\mu)\rho(\nu)
\left[1-\left({\sin x\over x}\right)^2\right] \eqno(1.13)
$$
$$
x=2\pi N\de\mu\rho(\bar\mu). \eqno(1.14)
$$
This result has long been known in the literature.
For $(\mu-\nu)$ small this implies that $\rho(\mu,\,\nu)$ vanishes as
$$
\rho (\mu,\nu) \sim {1\over3} \pi^2 \rho(\bar\mu)^4 [N(\mu-\nu)]^2. \eqno(1.15)
$$
as expected from the Van~der~Monde determinant in the measure.
\par
\item{}Note in this context that it is
 incorrect to replace $\sin^2x$ by its average
$1/2$, as is sometimes done in the literature.
\par
\item{(3.)}
The wild oscillation of $\rho_c(\mu,\nu)$ is entirely as expected since between
$\mu$ and $\nu$ finitely separated there are in general O(N) eigenvalues. Thus,
it is natural to smooth $\rho_c(\mu,\nu)$ by integrating over intervals
$\delta \mu$ and $\delta \nu$ large compared to O($N^{-1})$
 but small compared to
O($N^0$) centered around $\mu$ and $\nu$ respectively. We then
obtain\refto{BZ1}
$$
\rho^{{\rm smooth}}_c (\mu,\,\nu)  =
{-1\over2N^2\pi^2} {1\over(\mu-\nu)^2}
{(a^2-\mu\nu)\over[(a^2-\mu^2)(a^2-\nu^2)]^{1/2}}. \eqno(1.16)
$$
We find this result rather remarkable since the density $\rho(\mu)$ is
completely non-universal. The only dependence on $V$ appears through a.

\item{(4.)}
We have also computed the three- and four-point connected correlation
functions. When the oscillations in these functions are smoothed over we found
that they vanish identically to O($N^{-3})$ and O($N^{-4}$) respectively. We
conjectured that the smoothed $p$-point connected correlation function
similarly
vanishes to O($N^{-p}$).

\item{(5.)} We can show that the results in (1),(2),(3) hold for an ensemble
much more general than the one defined in (1.1) and (1.2), namely, an ensemble
defined with
$$
\eqalign{
H(\varphi)&={1\over N}Tr V(\varphi)+{1\over N^2}Tr W_1(\varphi)Tr W_2(\varphi)
\cr
&+{1\over N^3}Tr X_1(\varphi)Tr X_2(\varphi)Tr X_3(\varphi)
+\dots \cr}\eqno(1.17)
$$
with $W,X,Y,\dots$ \etc arbitrary polynomials. Indeed, it is easy to
see\refto{BZ2} that this ensemble can be further generalized by replacing, for
example, the third term in (1.13) by
$$
{1\over N^3} \sum_\a Tr X_1^\a (\varphi)Tr X_2^\a(\varphi)Tr X_3^\a(\varphi)
 \eqno(1.18)
$$
 with $X_i^\a$ a polynomial. This ensemble appears to us to be the most general
ensemble invariant under unitary transformations, \ie
$$
P(U^\dagger\varphi U)=P(\varphi) \eqno(1.19)
$$
except for some rather singular examples.

\item{(6.)} Wigner\refto{WIG} also studied non-unitary ensembles, for example
an ensemble of matrices $\varphi$ whose matrix elements take on the value
$\pm v /\sqrt{N}$ with equal probability. Using a renormalization
group\refto{BZJ}
 inspired approach we can show that $\rho(\mu)$ for this class
of matrix is universal and equal to the $\rho(\mu)$ for the simple Gaussian
ensemble  defined with $V(\varphi)={m^2\over2}\varphi^2$ in (1.2), namely
$$
\rho(\mu)={1\over\pi}\sqrt{a^2-\mu^2} \eqno(1.20)
$$
that is, Wigner's well-known semi-circle law.

\item{(7.)} In \Ref{BZ2}, we outline an argument showing that the results
stated
 in (1), (2),
(3), and (4) also hold for the non-unitary ensemble mentioned in (6).

Some of our results appear to overlap with results obtained in recent
literature.\refto{BEE}  In particular, our smoothed universal connected
correlation (1.16) appears to have been discovered also by Beenakker in an
interesting work,\refto{BEE1} although we have not yet seen a full derivation.

Given our correlation functions (1.11), (1.13), and (1.16), we can proceed to
determine the mean square fluctuation of physical quantities such as
conductance. For any quantity $A(\varphi)$ defined by the trace of some
function of $\varphi$,
the mean square fluctuation is clearly given by
$$
{\rm var\/}\ A\equiv\vev{A^2}-\vev{A}^2=\int d\mu d\nu \rho_c (\mu,\nu)
A(\mu)A(\nu)\eqno(1.21)
$$
We would like to emphasize that it is incorrect  to put for $\rho_c(\mu,\nu)$
the smoothed correlation $\rho_c^{{\rm smooth}}(\mu,\nu)\ \a\  {1\over
(\mu-\nu)^2}$
given in (1.16). The singularity in $\rho_c^{{\rm smooth}}(\mu,\nu)$ as
$\mu-\nu\to0$ would produce a divergent integral. Within our discussion, there
is of course no difficulty whatsoever, since $\rho_c^{{\rm smooth}} (\mu,\nu)$
was derived with the explicit proviso that $\mu-\nu$ is of  $0(N^0)$ and the
true $\rho_c(\mu,\nu)$ given in (1.11) is perfectly smooth as $\mu-\nu\to0$.
Calculation of the variance of the conductance given in the recent
literature\refto{BEE1} appears to us to involve simply replacing  $\rho_c$
by   $\rho_c^{{\rm
smooth}}$     without justification.

{}From the definitions for $\rho(\mu)$ and $\rho_c(\mu,\nu)$, it is easy to
derive the  response of $\rho(\mu)$ under a change in the potential $V(\nu)$,
as pointed out by Beenakker\refto{BEE1}:
$$
\de\rho(\mu)= -N^2\int d\nu\rho_c(\mu,\nu)\de V(\nu) \eqno(1.22)
$$
Again, it would be tempting to replace\refto{BEE1} $\rho_c(\mu,\nu)$ by
$\rho_c^{{\rm smooth}}(\mu,\nu)$. However, this illegitimate procedure would
lead to a divergent integral. The variation $\de\rho(\mu)/\de V(\nu)$ appears
to pose a rather tedious calculation with $P$ and $a$ in (1.10)
 depending on $V$ in
a complicated way.

We took $V$ to be an even polynomial for simplicity, so that
the density of eigenvalues is a symmetric function between the endpoints $-a$
 and $+a$.
It is a simple matter to shift the spectrum. Clearly, if we replace $\phi$
in $V$ by $\phi - d I$ (with $I$ the unit matrix and $d$ the shift)
the density of eigenvalues
would be non-zero  between $c = -a + d$ and $b = a + d$. The universal
correlation function is then trivially shifted to read
$$
\rho_c^{{\rm smooth}}(\mu\,\nu)=-{1\over2N^2\pi^2} {1\over(\mu-\nu)^2}
 {[-bc+{1\over2}(b+c)(\mu+\nu)-\mu\nu]\over
   [(b-\mu)(\mu-c)(b-\nu)(\nu-c)]^{1/2}} \eqno(1.23)
$$

In section II, we will develop a diagrammatic method which will enable us
to study
``time" dependent correlation between the eigenvalues, and which when the time
dependence is suppressed allows us to recover many of the results mentioned
above. In section III we compute the current-current correlation function. In
section IV we study a class of Hamiltonians consisting of the sum of a
deterministic term and a random term. The correlation between the eigenvalues
when the deterministic term is varied is calculated.

\head{II. ``TIME" DEPENDENT CORRELATION}

In this paper, we introduce a diagrammatic method to study the correlation
between eigenvalues for a time-dependent ensemble of hermitean matrices with a
probability distribution defined by
$$
P(\varphi)={1\over Z} \exp-\int_{-T}^T dt\  Tr \left[{1\over2}
({d\varphi\over dt})^2
 + V(\varphi)\right] \eqno(2.1)
$$
We take $T\to\infty$. Here the matrix $\varphi(t)$ depends on time, or more
generally, on some external parameter we are allowed to vary.
Physically, we may apply our results obtained below to disordered systems in
which the disorder may vary. In going from (1.1) to (2.1) we are moving from
zero-dimensional field theory (\ie an integral) to one-dimensional field theory
(\ie Euclidean quantum mechanics).
In the language of string theory, we move from a central charge $c=0$ theory to
a $c=1$ theory.
As a byproduct we show how some of our
previous results mentioned in Section~I may be recovered as a special case.
For ease of presentation, we will take $V(\varphi)={m^2\over2}\varphi^2$ to be
Gaussian and indicate below how our results may be generalized.

The one-point Green's function, or more generally the propagator,
$$
G_{ij}(z)\equiv\vev{({1\over z-\varphi(t)})_{ij}} \eqno(2.2)
$$
can be readily determined since due to time translation invariance it
does not in fact
 depend on time.  Using the usual Feynman diagram expansion, we find
immediately that in the large $N$ limit $G_{ij}(z)$ is given by planar Feynman
diagrams (generalized rainbow) as indicated in Fig.~(1a).

It is perhaps useful,  borrowing the terminology of large N QCD from the
particle physics literature,
to speak of the single line in Fig~(1a) as representing
 quark propagators and the double lines as gluon propagators.
The quark propagator is given simply by $1/z$ while the gluon propagator is
given by
$$
\eqalign{
D_{ij,kl} (t) & \equiv \vev{\varphi_{ij} (t)\varphi_{kl}(0)} \cr
& = \de_{il}\de_{jk}{1\over N}\int {d\om\over2\pi} {e^{i\om t}\over\om^2+m^2}
 =\de_{il}\de_{jk}{1\over 2Nm}e^{-m|t|}\cr}\eqno(2.3)
$$

We will now immediately generalize to arbitrary time dependence by replacing
(2.3) by
$$
D_{ij,kl} (t)
 = \de_{il}\de_{jk} {\si^2\over N}
e^{-u(t)}
\eqno(2.4)
$$
For instance, the time dependence in (2.1) may be changed in such a way that
the ``momentum space" propagator $\om^2+m^2$  in (2.4) is replaced by
${1\over\om^2+\ga|\om|+m^2}$ or ${1\over\om^4+\a\om^2+m^2}$ for example. In
other words, in (2.1) we can replace ${1\over2}
({d\varphi\over dt})^2$ by
$\varphi K({d\over dt} ) \varphi$ with K any reasonable
function. Our only requirement is that $u$ is a smooth function of $t$ and
 does not blow up as $t$ goes to zero.

Introducing as
usual the one-particle irreducible self energy
$\Si_{ij}(z)$, we can write the generalized rainbow integral equation
(Fig.~(1b))
$$
\Si(z)= \sigma^2 {1\over z-\Si(z)} = \sigma^2 G(z) \eqno(2.5)
$$
Here we have used the fact, immediately obvious from examining the Feynman
diagrams, that $\Si_{ij}(z)$ is equal to $\de_{ij}$ $\Sigma(z)$
  and the fact that the gluon
propagator only appears at equal time
$D_{ij,kl}(0)\equiv\de_{il}\de_{jk}{1\over N}{\sigma}^2$. Note
that the quark does
not know
about time. Solving the quadratic equation for $\Sigma$  we
obtain the Green's function as defined in (1.3)
$$
G(z) ={1\over2\si^2}(z -  \sqrt{z^2-4{\sigma^2}}) \eqno(2.6)
$$
Taking the absorptive part we recover immediately Wigner's semicircle law as
given in (1.20).

Incidentally, within this diagrammatic approach, band matrices can be treated
immediately. Let the matrices $\varphi$ be restricted so that $\varphi_{ij}$
vanishes unless $|i-j| < bN/ 2$ with $b <1$. Such matrices
 describe, for example, the hopping of a
single electron on a one-dimensional lattice with random hopping amplitudes.
The essential feature is that from each site the electron can hop to
$O(1/N)$ sites. Looking at the Feynman diagrams, we see
that in the generalized rainbow integral equation we simply restrict the range
of summation from $N$ to $bN$ and thus instead of (2.3) we obtain
$$
\Si(z)= {b\sigma}^2 {1\over z-\Si(z)} = {b\sigma}^2 G(z) \eqno(2.7)
$$
Thus, we have the same distribution of eigenvalues with a suitable
re-definition of the endpoints.

Let us now move on to the connected  2-point Green's function
$$
\eqalign{
G_c(z,w,t)&\equiv\vev{{1\over N^2}tr{1\over z-\varphi(t)}tr{1\over
w-\varphi(0)}
}_C
\cr
&={1\over N^2}\sum^\infty_{m=0}\sum^\infty_{n=0}{1\over z^{m+1}w^{n+1}}
\vev{tr\varphi^m(t)tr\varphi^n(0)}_C \cr}      \eqno(2.8)
$$
Henceforth for the sake of notational clarity we will set $\sigma$ to unity; it
can always be recovered by dimensional analysis.
Diagrammatically, the expression for
 $G_c(z,w,t)$ can be described as two separate quark loops, carrying
``momentum" $z$ and $w$ respectively, interacting by emitting
and absorbing gluons (see Fig.~(2a)).

With a Gaussian distribution for $\varphi$, we can readily ``Wick-contract"
the expression
$\vev{tr\varphi^m(t)tr\varphi^n(0)}_C$. Let us begin by ignoring contractions
within
the same trace (in which case $m$ and $n$ are required to be equal). In the
large $N$ limit, the dominant graphs
(see Fig.~(2b)) are given essentially by ``ladder graphs"
(with one crossing) which immediately sum to
$$
N^2 G_c(z,w) = {1\over(z w)^2}  {1\over(1-{1\over z w})^2} \eqno(2.9)
$$

We next include Wick-contractions within the same trace in $\vev{tr\varphi^m
tr\varphi^n}$. We see that graphically these contractions describe vortex
and self
energy corrections. The vortex corrections can be immediately summed: the
expression in (2.9) is to be multiplied by two factors, the factor
$$
{1\over(1-{1\over z^2})^2} \eqno(2.10)
$$
and a similar factor with $z$ replaced by $w$. Finally, according to our
earlier
discussion, self-energy corrections are included immediately by replacing the
bare quark propagator $1 / z$ by the dressed propagator $G(z)$ (and
similarly for $1 / w$ of course.) We obtain finally the remarkably compact
result
$$
N^2G_c(z,w)= {1\over(1-G(z)G(w))^2} \left[{G^2(z)\over1-G^2(z)}\right]
\left[{G^2(w)\over1-G^2(w)}\right]    \eqno(2.11)
$$

Finally, we have to put in the time dependence, which we have ignored so far.
We note first of all that the vertex and
self-energy corrections contain no time dependence, since there the gluons
always begin and end on the
same quark line. Thus, the time dependent 2-point
 connected Green's function can be
written down immediately as
$$
N^2G_c(z,w,t)= {e^{-u(t)}\over(1-e^{-u(t)}G(z)G(w))^2}
 \left[{G^2(z)\over1-G^2(z)}
\right]
\left[{G^2(w)\over1-G^2(w)}\right]
\eqno(2.12)
$$

To obtain the connected correlation function between energy
eigenvalues, we have to take
the double absorptive part of $G_c(z,w,t)$
 as indicated in (1.7). It is most convenient to introduce
angular variables: from (2.6) we see that we may write
$$
G(\mu+i\epsilon) = -i \eta e^{i \eta \theta} \eqno(2.13)
$$
where $\eta$ = the sign of $\epsilon$ and
$$
\sin \theta \equiv \mu / a.\eqno(2.14)
$$
 Similarly, we write
$$
G(\nu+i\delta) = -i \xi e^{i \xi \phi}     \eqno(2.15)
$$
with $\xi$ = the sign of $\delta$ and
$$
\sin \phi \equiv \nu / a.\eqno(2.16)
$$
 As $\mu$ and
$\nu$ vary over their
 allowed ranges, from $ -a$ to $+a$, $\theta$ and $\phi$ range vary
from $-\pi/ 2$ to $\pi/ 2$.

We can now readily compute (in the notation of (1.8))
$$
8 N^2 G_c(++)  = {1\over\cos \th\cos\phi}
\left\{{1 \over[1+{\rm ch\ u}\cos(\th+\phi)-i {\rm sh\ u}
 \sin (\theta+\phi)]}\right\}
\eqno(2.17)
$$
Proceeding in this way, we obtain one of our main results
$$
\eqalign{
& - 16\pi^2 N^2\rho_c(\mu,\,\nu) \cr
& = {1\over\cos \th\cos\phi}
\left\{{1+ {\rm ch\ u}\cos(\th+\phi)\over[{\rm ch\ u}+\cos(\th+\phi)]^2} +
(\phi\to-\phi+\pi)\right\} \cr} \eqno(2.18)
$$
Note that crossing symmetry $(\mu\leftrightarrow\nu,\, t\to-t)$ clearly holds.
  For
$t=0$, that is, $u=0$ (since any non-zero $u(0)$
 can be absorbed into $\sigma^2$),
 we recover immediately our previous result (1.16),
 obtained
 by the orthogonal polynomial method.
Note that time acts as a regulator for the singularity when $\mu = \nu$: for
time not equal to zero, we can set $\mu = \nu$ without difficulty and obtain
$$
8\pi^2N^2\rho_c(\mu,\mu)={1\over\cos^2\th} {1\over u^2} +\dots \eqno(2.19)
$$
For $\mu \neq \nu$ and small time $u << \theta - \phi$ we have
$$
 -16\pi^2 N^2\rho_c(\mu,\,\nu)={1\over\cos\th\cos\phi}
\left\{ {1\over1+\cos(\th+\phi)} \left[1-{1-{1\over2}\cos(\th+\phi) \over
       1+\cos(\th+\phi)}u^2\right] +(\phi\to-\phi+\pi) \right\}
\eqno(2.20)
$$

  In the long time limit, $u \rightarrow \infty$, we find
$$
4\pi^2N^2\rho_c(\mu,\,\nu)\to e^{-u}\tan\th\tan\phi \eqno(2.21)
$$

As expected the correlation vanishes exponentially in time. Notice however that
a memory of the spatial correlation is retained even at arbitrarily large time.

We thus conclude that the density-density correlation is universal in space for
all time, in the sense that it does not depend on V at all.

It is noteworthy that in the diagrammatic approach the correlation function is
``automatically" smoothed. An interchange of limits is responsible. Here in
computing $G(z,w)$ we are
taking $N$ to infinity and then letting $z$
approach the real axis. In our previous work\refto{BZ1} we use orthogonal
polynomials to calculate $\rho_c(\mu, \nu)$ directly. In effect, we
 let $z$ sit on
the real axis before taking $N$ to infinity and thus the discrete pole
structure
is visible. In this sense, the orthogonal polynomial method is more powerful
and informative.

Having now analyzed the Gaussian case we now discuss how our results could hold
more generally. We distinguish between the ``trace class" defined in
 (1.1) and the
``Wigner class" defined in (6).

 For the Wigner class,
let us focus on the example in which the probability of the distribution matrix
element $\varphi_{ij}$ is given by
$$
P(\varphi_{ij}) \propto e^{-N^2 {(|\varphi_{ij}|^2-{v^2\over N^2})^2 } }
\eqno(2.22)
$$
corresponding essentially to the example mentioned in item
(6) in Section~I (see \Ref{BZ2}). The
quartic interaction $\sim N^2|\varphi_{ij}|^4$
 would contribute to  Feynman diagrams such as the one in fig
(3a).
Counting powers, we see that this graph is of order
$N^{-4}NN^2=N^{-1}$ and so is
suppressed relative to the graphs in Fig.~(1a). Reasoning along this line, we
 see immediately
 that the distribution of eigenvalues is universal, a long-known result that we
also derived recently using a renormalization group inspired
approach.\refto{BZ2} As a
bonus, we obtain immediately in the present diagrammatic approach that the
correlation function is also universal. (Incidentally, this result is not at
all easy to obtain with the renormalization group approach of \Ref{BZ2}.)

For the trace class, the interaction, for example the quartic term in
$V(\varphi)$,  would generate
 Feynman diagrams such as the one in figure (3b). Counting  powers
of $N$ we see that this graph is in no way suppressed relative to the graphs in
Fig.~(1a). This is in fact a gratifying conclusion as we know from \Ref{BIPZ}
(See eq. (1.10)) that in the trace class, in sharp contrast to the Wigner
class,
the distribution of eigenvalues is in fact not universal. It appears to us that
within the diagrammatic approach, it would be rather involved to demonstrate
the universality of the correlation function for the trace class, namely the
result we obtained in \Ref{BZ1} using the method of orthogonal
polynomials. We would have to show that the effects of the arbitrary polynomial
interactions contained in $V$
 in  (1.1) can be summed up and absorbed completely in
the endpoint value $a$ of the spectrum.

We find it remarkable that in this subject results easily obtained in one
approach are apparently rather difficult to prove in another.

\head{III. Current-current Correlations}

As is well-known, the ensemble in (1.1) may be thought of as describing the
statistical mechanics of a gas in $1+0$ dimensional space time. The partition
function
$$
\eqalign{
Z &= \int d\varphi e^{-NtrV(\varphi)}\cr
&=C\int d\la_1\dots d\la_N e^{-N\sum_i V(\la_i)+\sum_{i<j}\log(\la_i-\la_j)^2}
\cr} \eqno(3.1)
$$
where $C$ is an irrelevant overall constant. The logarithmic term comes from
the
well-known Jacobian connecting $d\varphi$ to $d\la_1d\la_2\dots d\la_N$.
Regarding
$\la_i$ as the position of ``$i^{{\rm th}}$ particle" on the real line, we see
that (3.1) describes a one-dimensional gas of $N$ particles interacting with
each other via a logarithmic repulsion while confined by an external potential
$V(\la)$. Intuitively then, it becomes entirely clear that the density of the
gas $\rho(\mu)$ has no reason to be universal: it should certainly depend
 on $V$. It is less clear why the change in
density $\de \rho(\mu)$ at
 $\mu$ due to a change in the potential $\de V(\nu)$ at
$\nu$ should be universal, and indeed, this universality holds only when we
smooth over the discrete character of the gas.

The generalization in (2.1) then corresponds to allowing the particles to move
in a $1 + 1$ dimensional spacetime.
With the density operator defined by
$$
\rho (\mu, t) = {1\over N}\sum_i\de (\mu-\la_i(t)) \eqno(3.2)
$$
we clearly have the conservation law
$$
{\part \rho\over\part t}+{\part J\over\part\mu}=0  \eqno(3.3)
$$
with the current operator defined by
$$
J(\mu,t)={1\over N}\sum_i{d\la_i\over dt}\de (\mu-\la_i(t)) \eqno(3.4)
$$
Thus, we are led to study the current-current correlation function
$$
\vev{J(\mu,t)J(\nu,0)} = - {\part^2\over \part t^2} \int^\mu_{-a}d\mu'
\int^\nu_{-a}d\nu' \vev{\rho(\mu',\, t)\rho(\nu',0)}_c   \eqno(3.5)
$$
Note that $\vev{JJ}$ is connected by definition.

 The double integral in (3.5) may be explicitly evaluated.
 We find that in effect
$\rho_c(\mu,\nu)$ may be written as
$$
-4\pi^2N^2\rho_c(\mu,\,\nu)={\part\over\part\mu} {\part\over\part\nu} \log
\left[ {{\rm ch}\ u+\cos(\th+\phi)\over {\rm ch}\ u-\cos(\th-\phi)} \right]
\eqno(3.6)
$$
In particular, at equal time, we have
$$
-4\pi^2N^2\rho_c(\mu,\,\nu)={\part\over\part\mu} {\part\over\part\nu} \log
\left[{a^2-\mu\nu+\sqrt{(a^2-\mu^2)(a^2-\nu^2)}\over
a^2-\mu\nu-\sqrt{(a^2-\mu^2)(a^2-\nu^2)}} \right] \eqno(3.7)
$$
An equivalent form reads
$$
2\pi^2N^2\rho_c(\mu,\,\nu) = {\part\over\part\mu} {\part\over\part\nu} \log
\left[ {\sqrt{{a-\mu\over a+\mu}} - \sqrt{{a-\nu\over a+\nu}}\over
       \sqrt{{a-\mu\over a+\mu}} + \sqrt{{a-\nu\over a+\nu}}}\right]
\eqno(3.7a)
$$
The current-current correlation function then follows immediately
$$
\eqalign{
4\pi^2N^2 \langle J(\mu,\,t)J(\nu,\,0)\rangle & = {\part^2\over\part t^2} \log
\left[{{\rm ch}\ u+\cos(\th+\phi)\over {\rm ch}\ u-\cos((\th-\phi)}\right] \cr
 & = {{\rm sh\ u}\over {\rm ch}\ u +\cos(\th+\phi)} \ddot u +
{(1+{\rm ch}\ u\cos(\th+\phi)\over({\rm ch}\ u +\cos(\th+\phi))^2} \dot u^2 -
(\phi\to-\phi+\pi) \cr} \eqno(3.8)
$$

We have thus obtained the current-current correlation function for arbitrary
separation in space and time.

First, the dependence on space is
universal: as a function of $\theta$   and  $\phi$,  the current-current
correlation, just like the density-density correlation from which it is
derived, does not depend on the potential $V$.

The dependence on time, in contrast, is non-universal: clearly, the dependence
of   $u$  on  $t$  enters. In special cases, however, the specific functional
form may be seen to drop out. From (3.8), we see that at the same point in
space, that is when  $\theta  = \phi$, and $u$ small,
we have
$$
2\pi^2N^2 \langle J(\mu,\,t)J(\nu,\,0)\rangle  = {\part^2\over\part t^2}
\log u + \dots \eqno(3.9)
$$

Thus, if as $t\rightarrow 0$, $u$ vanishes like $u\rightarrow  \alpha
t^\gamma$, then
 $$
2\pi^2N^2 \langle JJ\rangle  \to -\gamma / t^2 \eqno(3.10)
$$
We
have universality in the sense that the unknown constant $\alpha$ has dropped
out. With the further assumption that $\gamma = 1$, which is reasonable but
certainly not required, we obtain the universal statement
$$
2\pi^2 N^2 \langle JJ\rangle \to -1/ t^2 \eqno(3.11)
$$

We also observe the curiosity that at  $ \theta =\phi =\pm \pi / 2$,
 the current-current
correlation vanishes identically for all time.

For large time, we obtain
$$
\pi ^2N^2\langle J(\mu,\,t)J(\nu,\,0)\rangle \to \cos \th \cos \phi
(\dot u^2-\ddot u) e^{-u}    \eqno(3.12)
$$
        This is of course not independent of (2.19).

We have learned that the universality of $\langle JJ \rangle$  at small time
has already been derived by Szafer and Altschuler\refto{SZA}
 and by Beenakker\refto{BEEX} using
apparently rather different methods and implicitly assuming that $\gamma=1$.
 To our knowledge, the complete form of
   $\langle JJ \rangle$ in (3.8) has not appeared before in the literature.

We can immediately generalize the preceding discussion to the case where many
external parameters $t_1,t_2,\dots,t_k\dots $
 are varied. This may be described picturesquely
as a many-time world in which the conservation laws
$$
{\part J^k\over\part\mu}={\part\rho\over\part t_k} \eqno(3.13)
$$
hold where the current with respect to time $t_k$ is defined by
$$
J^k(\mu)=\sum_i{\part\la_i\over\part t_k} \de(\mu-\la_i(t)) \eqno(3.14)
$$
The relation (3.5) immediately generalizes to
$$
\langle J^k(\mu,\,t)J^l(\nu,\,0)\rangle =-\part^k\part^l \int^\mu_{-a} d\mu'
\int^\nu_{-a} d\nu' \vev{\rho(\mu',\,t)\rho(\nu',\,0)}_c \eqno(3.15)
$$
where $\part^k\equiv {\part\over\part t_k}$.
 For the special case where $u$ is a function of $t =(\sum_kt^2_k)^{1/2}$,
 for example, we
obtain easily that
$$
2\pi^2N^2 \vev{J^k(\mu,\,t)J^l(\nu,\,0)} \to {t_kt_l\over(\Si_j
t_j^2)^{1/2}}\eqno(3.16)
$$
     for small time, with an assumption similar to the one that leads to
(3.11).

\head{ IV. DETERMINISTIC PLUS RANDOM}

Our diagrammatic approach allows us to study immediately the eigenvalues of a
Hamiltonian of the form $H=H_0+\varphi$  which consists of the sum of a
deterministic piece $H_0$ and a random piece $\varphi$    with a probability
distribution such as in (1.1). Pastur\refto{PAS} has
found the interesting relation that
$$
G(z) = G_0(z-G(z)) \eqno(4.1)
$$
where $G$ and $G_0$ are the Green's functions for $H$ and $H_0$ respectively.
We now
show that Pastur's relation follows immediately from our diagrammatic analysis.

Let us consider the Gaussian case and let $H_0$ be diagonal with diagonal
elements $\epsilon_i$.
Looking at the relevant Feynman diagrams, we see that we simply have to replace
the inverse quark propagator $ z$ by $z - \epsilon_i$ and thus obtain
$$
\Si(z)= {1\over N} \sum_k {1\over z-\eps_k-\Si(z)}=G(z) \eqno(4.2)
$$
We recognize this as (4.1).
The relation (4.1)
 while interestingly compact is not terribly useful in practice
as in solving for $G(z)$
 one would encounter a polynomial equation of degree
$ N+ 1$.

We will now demonstrate the power of the diagrammatic approach by showing how
we can immediately go beyond Pastur's relation and study correlation. Consider
the following class of physical problems. Suppose we change some external
parameters so that the deterministic Hamiltonian $H_0$ is changed to
$H_0'$. We would like to compute the correlation between the spectra of $H$
 and $H'=H_0'+\varphi$, in other words, we would like to compute
$$
\eqalign
G_c(z,w,H_0, H'_0)\equiv\vev{{1\over N} tr {1\over z
-H_0-\varphi}{1\over N} tr {1\over w-H_0'-\varphi}}_c \eqno(4.3)
$$
where the average, as before, is over the distribution of $\varphi$.

An example of this class of problems was recently studied by
Simons and Altschuler\refto{SIA}.    They considered the problem of a single
non-interacting electron moving in a ring threaded by a magnetic flux and with
the electron scattering on impurities in the ring. The magnetic flux
is then changed to some other value with $H$ changed
accordingly to $H'$. The correlation between the spectra of $H$ and $H'$
is apparently of great interest in the physics of mesoscopic systems.

We see immediately from Fig.~(2b)   that using the diagrammatic approach
we can determine
$G_c(z,w,H_0, H'_0)$
quite readily.
Let us define
$$
g_i(z)={1\over z-\eps_i-\Si(z)}  \eqno(4.4)
$$
where $\Si(z)$ is the solution of (4.2).
 We assume that $H'_0$ is also diagonal,
with diagonal elements $\eps'_i$. We define the analog of $g_i(z)$ for $H'_0$,
namely
$$
h_i(w) = {1\over w-\eps'_i-\Si'(w)} \eqno(4.5)
$$
where $\Si'(w)$ is the solution of the analog of
(4.2) for $H_0'$, namely (4.2) with
$\eps_i\to\eps'_i$,
$z\to w$,  and $\Si(z)\to\Si'(w)$. Let us also introduce the short hand
notation
$$
\eqalign{
g\cdot g & \equiv {1\over N} \Si_i g^2_i \cr
g\cdot h & \equiv {1\over N} \Si_i g_i h_i \cr
g^2\cdot h & \equiv {1\over N} \Si_i g_i^2 h_i \cr} \eqno(4.6)
$$
and so forth. Then
$$
N^2G_c(z,w,H_0,H'_0) = ({g^2\cdot h^2(1-g\cdot h)+(g^2\cdot h)(g\cdot h^2)\over
   (1-g\cdot h)^2}) {1\over1-g\cdot g} {1\over1-h\cdot h} \eqno(4.7)
$$
We see that this collapses to (2.11) when $H_0=H'_0=0$.

The number of physical situations covered by the result of this section is very
large. In particular, it should allow us to verify the intuitive expectations
concerning the universality of the correlations: the general belief is that
if the energy scale is such that the system can
explore the full extent of the disorder, one should recover the correlations
for the pure random
matrix correlations, irrespective of the non-random $H_0$. It is far from
obvious from the explicit representation (4.7). Clearly, our result (4.7) could
also be used to study questions such as localizations, or the influence of
white noise in various physical situations. Thus, we
 believe that the result in (4.7) would prove to be of
importance in studying many disordered systems.

\head{Appendix}

We have focussed on random hermitean matrices. In some physical situations, the
Hamiltonian matrix is in fact real symmetric. It is well-known in the
literature (see for example \Ref{PICH}) that going from the case of hermitean
random matrices to the case of real symmetric matrices we simply insert in
various formulas appropriate factors of 2. This is most easily seen by looking
at (3.1): for $\varphi$ real symmetric matrices we would have a factor of 1/2
in front of the logarithmic repulsion. This comes about because the Jacobian
connecting $d \varphi$ to $d\lambda_1
 d\lambda_2\dots d\lambda_N$ for the hermitean
case is the positive square root of the corresponding Jacobian for the real
symmetric case.

Let us now sketch exceedingly briefly how this factor of 2 emerges in the
diagrammatic
approach. Typically, we encounter
$\vev{tr \varphi^n tr \varphi^n}$. To indicate how
the argument goes, let us consider only Wick-contractions between the two
traces. After the first contraction, we have
$$
n \vev{\varphi_{ij}\varphi_{\alpha\beta}}\vev{(\varphi^{n-1})_{ji}
(\varphi^{n-1})_{\beta\alpha}}
=        n (\de_{i\alpha}\de_{j\beta} + \de_{i\beta}\de_{j\alpha})
\vev{(\varphi^{n-1})_{ji}(\varphi^{n-1})_{\beta\alpha}}
= 2 n \de_{i\beta}\de_{j\alpha}  \vev{\varphi^{n-1}_{ji}
\varphi^{n-1}_{\beta\alpha}}       \eqno(A.1)
$$
The last line follows from the fact that $\varphi$ is symmetric.  Proceeding,
we find easily that the above is equal to $2 n N^n$.
Note that for hermitean matrices the expression in the parenthesis would read
$\de_{i\alpha}\de_{j\beta}$ instead, and this accounts for the factor of 2
alluded to above.

Graphically, the fact that  $\vev{\varphi_{ij}\varphi_{\alpha\beta}}$
is equal to $ \de_{i\alpha}\de_{j\beta}
+ \de_{i\beta}\de_{j\alpha} $     rather than $\de_{i\alpha}\de_{j\beta}$
means that the double lines in the gluon propagator in the Feynman diagrams can
be twisted. The direct of counting of graphs becomes considerably more
involved.

We also note that the method of orthogonal polynomials used in \Ref{BZ1}
becomes rather complicated when we deal with real symmetric matrices.

\head{Acknowledgement}

While we were completing this work, we received from C.~Beenakker a helpful
correspondence calling our attention to his recent work, for which we are most
grateful. We would like to thank A.~Devoto for his warm hospitality allowing us
to complete this work on the beach of Chia. One of us (EB) thanks C. ~Beenakker
and Y.~Imry for a discussion. The other (AZ) thanks X.G.~Wen for a stimulating
discussion at Nordita, whose hospitality, as well as the hospitality of the
\'Ecole Normale Sup\'erieure, where this work was initiated, are gratefully
acknowledged. We thank A.~Felce for drawing the figures.
 This work was supported in part by the National Science
Foundation under Grant No. PHY89-04035 and the Institut Universitaire de
France.

\head{Figure Captions}

Fig (1.a) Feynman diagram expansion for the Green's function $G(z)$.

Fig (1.b) The generalized rainbow equation for the one-particle-irreducible
self energy $\Sigma(z)$.

Fig (2.a) Feynman diagram expansion for the connected two-point Green's
function. The dotted lines here represent the ``gluon" propagator.

Fig (2.b) Some typical low order Feynman diagrams contributing to the connected
two-point Green's function.

Fig (3) Non-Gaussian corrections to the propagator in the (a) Wigner class
and in the (b) trace class.

\references

\refis{BZ1} E. Br\'ezin and A. Zee, \np 402(FS), 613, 1993.

\refis{BZ2} E. Br\'ezin and A. Zee, {\sl Compt.\ Rend.\ Acad.\ Sci.\/}, to be
published 1993.

\refis{WIG} E. Wigner, {\sl Can.\ Math.\ Congr.\ Proc.\/} p.174 (University of
Toronto Press) and other papers reprinted in Porter, op. cit.

\refis{POR} C.E. Porter, {\it Statistical\ Theories\ of\  Spectra:\ \
Fluctuations\/}
(Academic Press, New York, 1965).

\refis{MEH} M.L. Mehta, {\it Random\ Matrices\/} (Academic Press, New York,
1991).

\refis{JER} See for instance, the Proceedings of the Eighth Jerusalem Winer
School, Two Dimensional Quantum Gravity and Random Surfaces, D.J. Gross and T.
Piran (Wold Scientific, Singapore, 1992).

\refis{BZJ} E. Br\'ezin and J. Zinn-Justin, \pl B288, 54, 1992.

\refis{BEE} C.W.J. Beenakker and B. Rejaei, Institut
Lorentz preprint (1993); R.A. Jalabert, J.-L. Pichard, and C.W.J. Beenakker,
Institut Lorentz preprint (1993).

\refis{SIA} B.D. Simons and B.L. Altschuler, \prl 70, 4063, 1993.

\refis{BEEX} C.W.J. Beenakker, \prl 70, 4126, 1993.

\refis{SZA} A. Szafer and B.L. Altschuler \prl 70, 587, 1993.

\refis{BEE1} C.W.J. Beenakker, \prl 70, 1155, 1993.

\refis{BEE2} C.W.J. Beenakker, \pr B47, 15763, 1993.

\refis{PICH} J.-L. Pichard, in {\it Quantum\ Coherence\ in\ Mesoscopic\ \
Systems\/} (NATO ASI series, Plenum, New York 1991) ed. by B. Kramer, we
thank Y. Meir for informing us of this reference.

\refis{BIPZ} E. Br\'ezin, C. Itzykson, G. Paris, and J.B. Zuber,
\journal  Comm. Math. Phys., 59, 35, 1978.

\refis{PAS} L.A. Pastur, \journal Theo. Math. Phys., 10, 67, 1972.

\endreferences

\end